\documentclass[superscriptaddress,amsmath,amssymb,aps,prl,reprint]{revtex4-1}

\usepackage{graphicx}
\usepackage{dcolumn}
\usepackage{bm}
\usepackage{hyperref}
\usepackage{notes2bib}
\usepackage{siunitx}
\usepackage{amsmath}
\usepackage[usenames,dvipsnames]{xcolor}
\usepackage[utf8]{inputenc}

\setcitestyle{super}

\newcommand{\Vgone}{\ensuremath V_{\mathrm{G1}}}
\newcommand{\Vgtwo}{\ensuremath V_{\mathrm{G2}}}
\newcommand{\Ic}{\ensuremath I_\mathrm{c}}
\newcommand{\Icone}{\ensuremath I_\mathrm{c,1}}
\newcommand{\Ictwo}{\ensuremath I_\mathrm{c,2}}
\newcommand{\none}{\ensuremath n_\mathrm{1}}
\newcommand{\ntwo}{\ensuremath n_\mathrm{2}}

\begin{document}

\title{A Tunable Monolithic SQUID in Twisted Bilayer Graphene}

\author{El\'ias Portolés}	
\email{eliaspo@phys.ethz.ch}
\author{Shuichi Iwakiri}
\author{Giulia Zheng}
\author{Peter Rickhaus}
\affiliation{Solid State Physics Laboratory, ETH Zurich,~CH-8093~Zurich, Switzerland}
\author{Takashi Taniguchi}
\affiliation{International Center for Materials Nanoarchitectonics,
National Institute for Materials Science,  1-1 Namiki, Tsukuba 305-0044, Japan}
\author{Kenji Watanabe}
\affiliation{Research Center for Functional Materials,
National Institute for Materials Science, 1-1 Namiki, Tsukuba 305-0044, Japan}
\author{Thomas Ihn}
\author{Klaus Ensslin}
\affiliation{Solid State Physics Laboratory, ETH Zurich,~CH-8093~Zurich, Switzerland}
\affiliation{Quantum Center, ETH Zurich,~CH-8093 Zurich, Switzerland}
\author{Folkert K. de Vries}
\affiliation{Solid State Physics Laboratory, ETH Zurich,~CH-8093~Zurich, Switzerland}


\maketitle

\textbf{
Magic-angle twisted bilayer graphene (MATBG) hosts a number of correlated states of matter that can be tuned by electrostatic doping~\cite{Cao2018_1, Cao2018_2, Lu2019, Li2010}. Superconductivity has drawn considerable attention and the mechanism behind it is a topic of active discussion~\cite{Khalaf2021}. MATBG has been experimentally characterized by numerous transport~\cite{Yankowitz2019, Das2021} and scanning-probe~\cite{Jiang2019, Nuckolls2020, Choi2021} experiments.
The material has also emerged as a versatile platform for superconducting electronics, as proven by the realization of monolithic Josephson junctions~\cite{deVries2021, Rodan-Legrain2021}.
However, even though phase-coherent phenomena have been measured, no control of the superconducting phase has been demonstrated so far.
Here, we present a Superconducting Quantum Interference Device (SQUID) in MATBG, where the superconducting phase difference is controlled through the magnetic field. 
We observe magneto-oscillations of the critical current, demonstrating long-range coherence agreeing with an effective charge of $2e$ for the superconducting charge carriers. 
We tune to both asymmetric and symmetric SQUID configurations by electrostatically controlling the critical currents through  the junctions. With this tunability, we study the inductances in the device, finding values of up to $\SI{2}{\micro H}$. 
Furthermore, we directly observe the current-phase relation of one of the Josephson junctions of the device.
Our results show that superconducting devices in MATBG can be scaled up and used to reveal properties of the material.
We expect this to foster a more systematic realization of devices of this type, increasing the accuracy with which microscopic characteristics of the material are extracted. We also envision more complex devices to emerge, such as phase-slip junctions or high kinetic inductance detectors.
}
\newline
\twocolumngrid

The recent realization of electrostatically defined Josephson junctions (JJs) in MATBG, witnessed by the observation of magnetic interference~\cite{Rodan-Legrain2021} and Shapiro steps~\cite{deVries2021}, has sparked interest in combining such devices with other states like orbital magnets~\cite{Diez-Merida2021}.
Gate-defined quantum devices are thus emerging as a probe of the physics hosted by MATBG that is complementary to bulk measurements or scanning-probe experiments.
When combining two JJs into a SQUID, the interference between the supercurrents through each arm as a function of an out-of-plane magnetic field leads to an oscillatory critical current across the device~\cite{Tinkham2004}.
Among other characteristics, the effective charge of the carriers, the system's inductance, or the current-phase relation (CPR) of the JJs can be estimated from such oscillations~\cite{Clarke2006}.
Beyond being a material probe, SQUIDs have an established history of applications as, for example, magnetometers~\cite{Clarke2006, Tinkham2004}, photon detectors~\cite{Day2003}, tunable resonators~\cite{Palacios-Laloy2008} or superconducting quantum bits~\cite{Koch2007, Orlando1999}.
An all-electrically-controllable monolithic SQUID would allow for \textit{in situ} tuning of its macroscopic parameters, offering new possibilities for controlling devices like highly balanced SQUIDs~\cite{Kemppinen2008} or frequency-tunable quantum bits \cite{Larsen2015}.

\begin{figure*}[t!]
\includegraphics[width=1\textwidth]{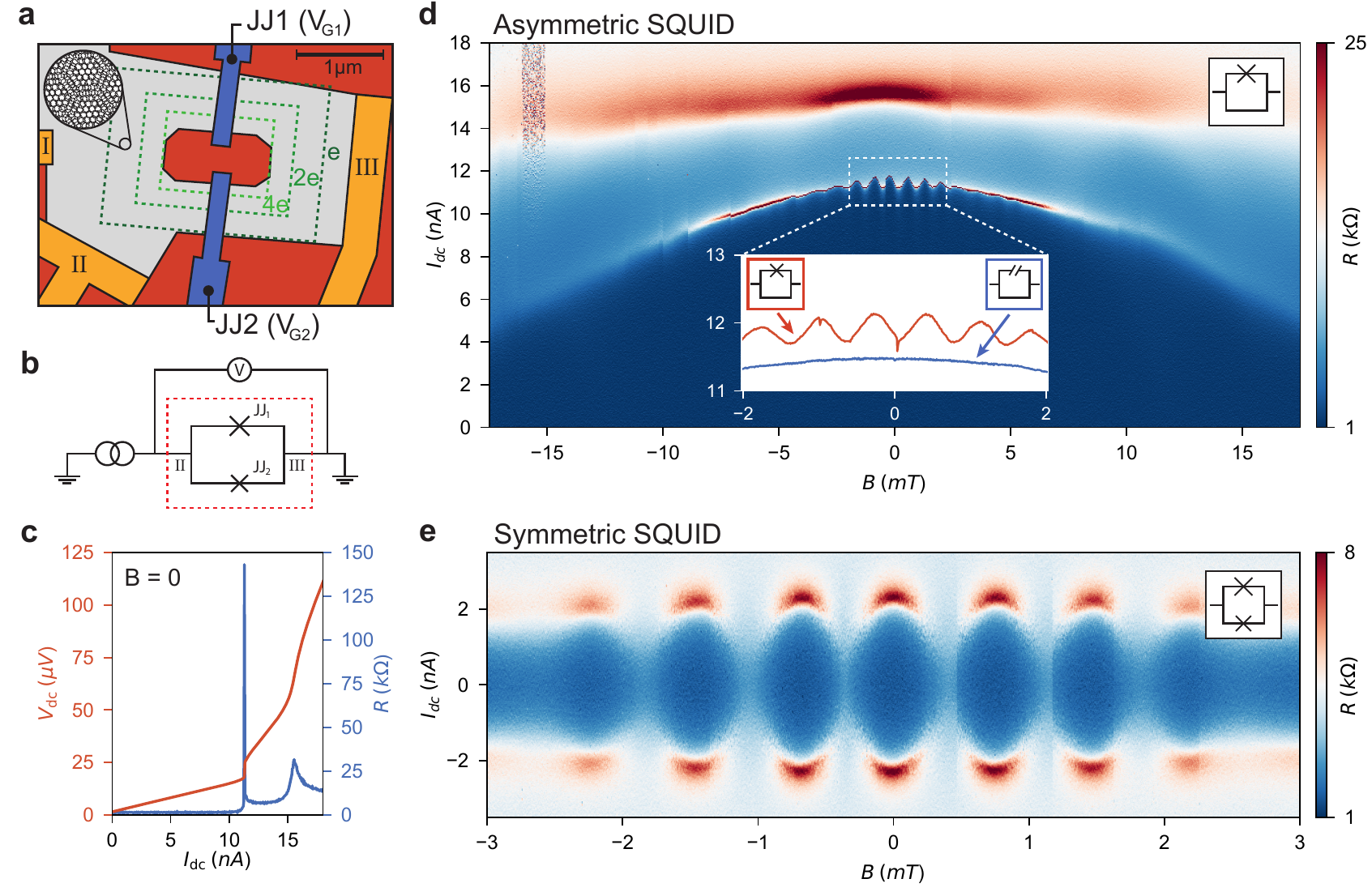}
\caption{\textbf{Tunable superconducting quantum interference device.
a} Device schematics to scale. Contacts are in yellow, top gates in blue, etched areas in red and MATBG in grey. The dashed green rectangles represent the different areas, being $S_{\mathrm{eff}} = \SI{3.22}{\micro m^2}$, $S_{\mathrm{eff}} = \SI{1.61}{\micro m^2}$ and $S_{\mathrm{eff}} = \SI{0.805}{\micro m^2}$ corresponding to an effective charge of $e$, $2e$ and $4e$ for the superconducting charge carriers, respectively. \textbf{b} Circuit equivalent representation of the device, hosting JJ1 and JJ2. The current bias, voltage measure setup is displayed. \textbf{c} Line trace (red) at zero magnetic field of the voltage drop across the device as a function of current bias $I_{\mathrm{dc}}$. The the contact resistance, present in the measurement due to the 2 terminal measurement setup, accounts for the finite slope in the superconducting regime. The differential resistance $dV/dI_{\mathrm{dc}}$ corresponding to the voltage drop is shown in blue. \textbf{d} Resistance across the device as a function of $I_{\mathrm{dc}}$ and out-of-plane magnetic field, with both arms being tuned to host their maximal supercurrent, showing oscillations of period $\Delta B = \SI{0.8}{mT}$ in the critical current. Dark blue corresponds to a superconducting path connecting both contacts, light blue to a state in which a part of the device not connected to both contacts is superconducting and red to the whole sample being normal. We subtract an offset in magnetic field (see Methods). The inset shows a line trace of the critical current in the region where oscillations occur when both arms of the SQUID are superconducting (red) or when only one is (blue). \textbf{e} Resistance across the device as a function of current bias and out-of-plane magnetic field when the SQUID is tuned to a symmetric configuration.}
\label{fig:1}
\end{figure*}

To date, evidence of superconducting interference in MATBG has been Fraunhofer-like patterns in single junction geometries~\cite{Rodan-Legrain2021, deVries2021}.
Here, we build on previous gate-defined JJ realizations and form a SQUID, a ring geometry with one JJ in each arm. 
To observe interference, a superconducting path must encircle the ring.
Fulfilling such a condition in MATBG remains a fabrication challenge because of twist angle inhomogeneity~\cite{Uri2020}, making our device a major leap forward in terms of complexity.
We observe coherent behavior over a distance at least an order of magnitude larger than in single junction interference~\cite{Rodan-Legrain2021, deVries2021}.
Finally, a SQUID grants experimental control of the superconducting phase, a knob upon which our study builds.

The device is depicted in Fig.~\ref{fig:1}a. 
The MATBG is encapsulated in hexagonal boron nitride and the twist angle averaged over the device is $0.95^\circ \pm 0.04^\circ$. 
A MATBG ring is defined by etching a hole in the center of the device, and contacted by three gold electrodes (I, II, III).
The data shown below are taken between contacts II and III (see Extended Data Figs.~\ref{fig:S_G}, \ref{fig:S_I} and \ref{fig:S_J} for other contact combinations).
Two local top gates, $\mathrm{G_1}$ and $\mathrm{G_2}$, control the local carrier densities $\none$ and $\ntwo$ in a $\SI{120}{nm}$ long region in arm 1 and arm 2 of the ring and a graphite back gate tunes the global carrier density $n$. 
We send a current from contact II to III, and measure the voltage drop between these contacts to calculate the two-terminal differential resistance $R = dV/dI_\mathrm{dc}$.
The circuit equivalent to the electrical setup is shown in Fig.~\ref{fig:1}b, where the dashed red box represents the device.
All measurements are performed in a $^3$He--$^4$He dilution refrigerator at a temperature of $\SI{50}{mK}$ unless stated otherwise.
Further details of the fabrication process, twist angle determination and measurement setup are given in the Methods section and Extended Data Figs.~\ref{fig:S_C} and \ref{fig:S_D}. 

We first find the optimal gate configuration at which the critical current $I_\mathrm{c}$ is maximized in each junction.
By applying a gate voltage, we tune the electron density, either of the whole device (in the case of the global back gate), or of the region underneath a gate (for the local top ones).
MATBG presents flat bands at densities above and below the charge neutrality point (CNP), where a filling factor $\nu$ is defined as the number of electrons per moiré unit cell~\cite{Cao2018_2, Lu2019}.
The most prominent superconducting region is expected for $-3 < \nu < -2$.
We measure $I_\mathrm{c}$ as a function of the global back gate voltage with zero volts applied to the top gates and extract, in such filling factor range, the density at which the critical current is highest $n_\mathrm{opt} = \SI{-1.27e12}{cm^{-2}}$ (see Extended Data Fig.~\ref{fig:S_E}). 
At this global density, we tune the top gate of JJ1 into the normal state and maximize the critical current of junction 2 using $V_\mathrm{G2}$ and vice versa (see Extended Data Fig.~\ref{fig:S_F}). 
We find the maximum critical currents of the two arms to be very different in magnitude, namely $\Icone=\SI{0.5}{nA}$ for JJ1, and $\Ictwo=\SI{11}{nA}$ for JJ2.

We observe a finite slope in voltage in the superconducting regime, due to the two-terminal measurement setup.
The pronounced peak in resistance at $I_{\mathrm{dc}} = \SI{11.5}{nA}$ indicates the critical current of the device.
The second peak at $I_{\mathrm{dc}} = \SI{15.5}{nA}$ most likely corresponds to a spatial region of higher critical current that does not extend all the way between the two contacts. 
We attribute this observation to twist angle inhomogeneity.

In order to demonstrate phase-coherence around the ring, we show the differential resistance as a function of $I_\mathrm{dc}$ and perpendicular magnetic field $B$ in Fig.~\ref{fig:1}d. 
We observe oscillations in $\Ic$ with period $\Delta B = \SI{0.8}{mT}$ and an amplitude of approximately $\SI{0.5}{nA}$ on top of a critical current background of maximally $\SI{11.5}{nA}$. 
When electrostatically tuning JJ1 out of the superconducting state, we suppress the oscillations, as seen in the inset of Fig.~\ref{fig:1}d.
Since $\Ictwo$ is smooth in $B$ and no Fraunhofer-like interference is seen when JJ1 is tuned to the normal state (See Extended Data Fig.~\ref{fig:S_B}), arm 2 is in a bulk superconducting state at this optimal $\ntwo$, indicated by the absence of the junction in the schematics in Fig.~\ref{fig:1}d.
The fact that oscillations only appear when both arms are superconducting, as shown in the inset of Fig.~\ref{fig:1}d, confirms that we have phase coherence around the entire ring.

The period of the oscillations is a direct measure the effective area of the loop~\cite{Clarke2006} $S_{\mathrm{eff}} = h/ (e^*\Delta B)$, with $e^*$ being the effective charge of the superconducting charge carriers. 
In Fig.~\ref{fig:1}a we plot rectangles corresponding to this area for different effective charges of the carriers and observe that the two extreme cases of $e^* = e$ and $e^* = 4e$ have an effective surface that barely fits the device, while $e^* = 2e$ leads to an interference path at the center of each arm. 
Therefore, the observed periodicity points towards the effective charge being $2e$.
This is consistent with the Shapiro steps observed in a single JJ geometry, a quantity which also depends on the charge of the carriers~\cite{deVries2021}.

For the oscillations shown in the inset of Fig.~\ref{fig:1}d, we observe switches at $B = \SI{-1}{mT}$ and $B = \SI{0}{mT}$. 
We cannot rule out that they are caused by our experimental setup and therefore only address them in the methods section and in Extended Data Fig.~\ref{fig:S_H}.

\begin{figure*}[t!]
\centering
\includegraphics[width=1\textwidth]{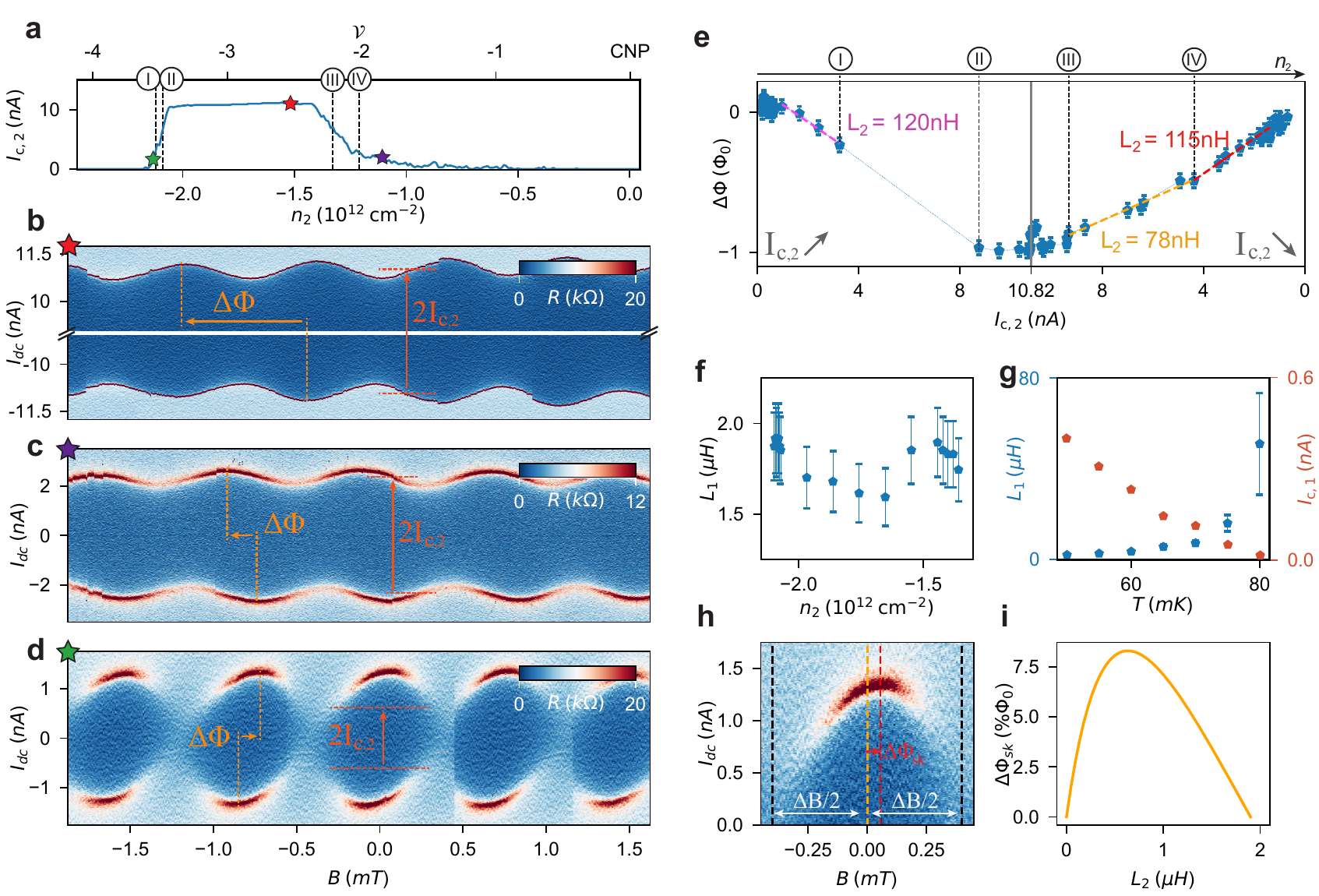}
\caption{\textbf{Tuning of the asymmetry and extraction of inductances. a} Critical current across arm 2 as a function of $\ntwo$ and the corresponding filling factor $\nu$. \textbf{b-d} Magnetic interference patterns taken at values of $\ntwo$ indicated in \textbf{a} by red, purple and green stars respectively. \textbf{e} Flux shift as a function of critical current in arm 2. The critical current axis is mirrored at its maximal value, $\Ictwo = \SI{10.82}{nA}$ while the top axis highlights the unevenly spaced, non-mirrored, density coordinates of particular points, corresponding to I-IV in \textbf{a}. Dashed lines correspond to local least-squares fits to the data and are accompanied by the resulting inductance values. \textbf{f} Inductance of arm 1 as a function of $\ntwo$ in the density region between II and III. \textbf{g} Critical current of arm 1 (red) and corresponding inductance $L_1$ (blue) as a function of temperature. \textbf{h} Zoom-in of \textbf{d}. The yellow dashed lines highlight the middle of the lobe, equidistant to each minimum (black dashed lines). The red dashed line highlight the positive field value at which the maximal critical current is attained. \textbf{i} Flux shift due to inductance asymmetry effects as a function of $L_2$ at a fixed value $L_1 = \SI{2}{\micro H}$.}
\label{fig:2}
\end{figure*}

In addition to phase coherence throughout the superconducting ring, a SQUID requires the presence of a JJ in each of its arms.
If both junctions host the same supercurrent, the total critical current across the device reaches zero periodically in magnetic field.
We leave $\none$ in its optimal position, tune $\ntwo$ so that $\Icone = \Ictwo$ and observe the periodic critical current modulation that goes close to zero shown in Fig.~\ref{fig:1}e.
The parabolic shape of the pattern is another characteristic of a symmetric SQUID~\cite{Clarke2006}, and further confirms that we do have a JJ in each arm, as depicted by the inset schematic in the figure.
In this electrostatic configuration our device is therefore a SQUID in its \textit{symmetric regime}.

We now turn to the all-electrical \textit{in situ} tunability of our device and its limits, as this can be a key asset in simplifying current architectures~\cite{Kemppinen2008, Wang2019}.
Having established that we have a SQUID, we interpret the data shown in Fig.~\ref{fig:1}d as arising from the most \textit{asymmetric regime} we can tune our SQUID into, where arm 2 hosts no JJ.
Finally, we are also able to locally turn off superconductivity in each arm by biasing the local gates further away from the superconducting dome (see Extended Data Fig.~\ref{fig:S_E}).
Therefore, we can tune arm 2 to having no superconductivity, a JJ or being a bulk superconductor, while for arm 1 we can turn off superconductivity or tune it to a JJ. 

After considering the limiting scenarios, we study the continuous evolution in asymmetry by changing $\ntwo$.
Figure~\ref{fig:2}a reveals the dependence of the critical current on the electron density $\ntwo$.
It shows a plateau at $\SI{11}{nA}$ between filling factors $\nu = -4$ and $\nu = -2$ and decreases sharply towards full filling of the lower flat band and more progressively towards the CNP. 
Similar behaviour has been observed before~\cite{deVries2021,Rodan-Legrain2021}. 
We keep $\Icone = \SI{0.5}{nA}$ constant and  in Fig.~\ref{fig:2}a indicate the points with stars where we fix $\ntwo$ for the magnetic interference patterns shown in Fig.~\ref{fig:2}b-d.
To avoid hysteresis due to heating~\cite{Courtois2008}, at each magnetic field value we take a trace from zero current bias towards both $I > 0$ and $I < 0$ (see Supplementary Information).
Figure~\ref{fig:2}b shows oscillations in the most asymmetric regime (same regime as in Fig.~\ref{fig:1}d). 
We observe a flux shift $\Delta \Phi$ in applied magnetic field and an offset $2\Ictwo$ between the averages of the positive and negative switching current. 
When reducing $\Ictwo$, we observe a less asymmetric regime (Fig.~\ref{fig:2}c), where we still have $\Icone < \Ictwo$. 
Finally, as we keep decreasing $\Ictwo$, we observe a fully symmetric situation, as previously shown in Fig.~\ref{fig:1}e, and even reach $\Icone > \Ictwo$ in Fig.~\ref{fig:2}d.

Because in 2D materials the cross section $\mathcal{A}$ is strongly reduced in the vertical direction compared to metallic superconductors, the kinetic inductance~\cite{Cardwell2002} $L_\mathrm{K} \propto 1/\mathcal{A}$ is expected to be orders of magnitude higher than for 3D devices.
This would open the door for phase-slip devices~\cite{Mooij2006} or improved high-kinetic-inductance photon detectors~\cite{Day2003}.
We now explain how one can extract inductances of each arm in our SQUID from the flux shift observed at different gate configurations.
The total flux threading a superconducting loop is given by $\Phi_\mathrm{T} = \Phi_\mathrm{a} + \Phi_\mathrm{s}$, where $\Phi_\mathrm{a}$ and $\Phi_\mathrm{s}$ are respectively the externally applied and the so-called self flux~\cite{Clarke2006}.
The latter originates from the screening current circulating in the SQUID loop and results in an offset in flux for negative and positive currents given by~\cite{Clarke2006} 
\begin{equation}
    \Delta \Phi = 2(L_\mathrm{1}I_\mathrm{1} - L_\mathrm{2}I_\mathrm{2}),
    \label{eq:inductance}
\end{equation}
where $L_\mathrm{1,2}$ is the inductance of the respective arm.
At the maximum switching current of the interference pattern, both junctions host their critical currents. 
Combining this with the previous relation, we deduce 
\begin{equation}
    L_\mathrm{2} = -(1/2)\frac{\partial (\Delta \Phi)}{\partial (\Ictwo)},
    \label{eq:single_inductance}
\end{equation} 
allowing us to extract the inductance of arm 2 from the flux offset evolution as a function $\ntwo$.
Equation~\eqref{eq:single_inductance} holds, however, only when $L_2$ is constant, thus can only be locally applied when $\Delta \Phi (\Ictwo)$ is linear.
It must also be noted that, at finite temperature, the current at which the junction switches out of the superconducting state (switching current) is different from the one at which it would do so at zero temperature (critical current).
If this difference is significant, this would result in a drift of the phase at the switching current of our junctions as we tune the local density.
This would subsequently lead to a contribution to the offset in field in the interference patterns.
However, based on switching statistics measurements performed in a device hosting a similar JJ~\cite{deVries2021}, we neglect this effect in our case (see Supplementary Information).

In Fig.~\ref{fig:2}e, we first analyze the evolution of the flux shift $\Delta \Phi$ as a function of the critical current in JJ2 while increasing the density.
For the sake of clarity, the critical current axis is mirrored at the maximal critical current. 
We divide the density into ranges defined by the points I to IV in Fig.~\ref{fig:2}a. 
Using Eq.~\eqref{eq:single_inductance}, we extract an estimate $L_2 = \SI{120}{nH}$ for values of the critical current up to point I (pink dashed line).
The region between indicated points I and II has no data due to the atypical interference patterns observed in that range of parameters, from which we cannot extract the inductance. 
Subsequently, in the region between point II and III, $\Delta \Phi$ stays relatively constant, resulting in a vanishing $L_2$.
Following the evolution of $\Delta \Phi$ from point III to IV and beyond, we observe a slight non-linearity at IV, corresponding to $\nu = -2$, as highlighted by the orange and red dashed lines in Fig.~\ref{fig:2}e, which correspond to $L_2 = \SI{78}{nH}$ and $L_2 = \SI{115}{nH}$ respectively.
Generally, by tuning the electron density through the band, we not only tune the critical current but also the inductance of arm 2.

The vanishing $L_2$ between II and III gives access to $L_1$ through Eq.~\eqref{eq:inductance} and using $L_2 \sim 0$. 
Figure~\ref{fig:2}f shows $L_1$ calculated by simplifying Eq.~\eqref{eq:inductance} to $L_1 = \Delta \Phi_\mathrm{s}/2I_1$, as $\ntwo$ is tuned.
We extract an inductance at base temperature (where it is expected to be lowest~\cite{Goswami2016}) $L_1$ of the order of $\SI{2}{\micro H}$, orders of magnitude higher than in other 2D superconductors~\cite{Goswami2016} or narrow 3D SQUID geometries~\cite{Nichele2020}.
The geometric contribution to the inductance $L$ is negligible compared to the measured values, confirming the inductance of the device to be mainly of kinetic origin (see Methods). 
Our result shows that a value of $\SI{2}{\micro H}$ can be reached while still having coherent transport, highlighting the potential of MATBG as a host for high inductance devices.

The evolution of the kinetic inductance with temperature gives access to that of microscopic parameters of the material such as the superfluid density or the effective mass of the superconducting charge carriers.
While staying in the same electrostatic configuration, we follow the evolution of $\Delta \Phi_\mathrm{s}$ and $\Icone$ as we increase the temperature until no oscillation is observed in the critical current (Fig.~\ref{fig:2}g).
We stress that the last point where a measurable oscillation is observed ($T = \SI{80}{mK}$) presents an amplitude of $\SI{15}{pA}$, a value which would be impossible to distinguish from noise without an interferometric measurement.
The kinetic inductance and critical currents are connected to microscopic parameters of the superconductor through $L_\mathrm{K} \propto m^*/(n_\mathrm{s}\mathcal{A})$ and $I_\mathrm{c} \propto n_\mathrm{s}$, with $m^*$ and $n_\mathrm{s}$ the effective mass and superfluid density, respectively~\cite{Tinkham2004}.
As the temperature increases, $\Delta \Phi_\mathrm{s} = L_1 \times \Icone$, independent of $n_\mathrm{s}$, remains constant while $\Icone$ drops, indicating an increase of the inductance, as depicted in Fig.~\ref{fig:2}g.
This suggests that we tune the local superfluid density with temperature, rather than the effective mass, the former vanishing as the critical temperature of arm 1 is reached.
This density vanishes as the critical temperature of arm 1 is approached, leading to the divergence of the kinetic inductance.
We measure values of up to $\SI{50}{\micro H}$ within the high experimental accuracy that the SQUID device allows for.

Due to the inductance asymmetry, the magnetic interference patterns can present a skewed shape~\cite{Tesche1977}.
This is the case for Fig.~\ref{fig:2}d, of which we show a zoom in Fig.~\ref{fig:2}h. 
We estimate the flux shift $\Delta \Phi_{\mathrm{sk}}$ due to the skewness to be of the order of $5\%$ of a flux quantum\footnote{The magnetic interference patterns where we see significant skewness, which happens only at $I_\mathrm{c} < \SI{1}{nA}$, are discarded when plotting Fig.~\ref{fig:2}e}.
By fixing $L_1 = \SI{2}{\micro H}$, we calculate this shift as a function of $L_2$, as shown in Fig.~\ref{fig:2}i, and obtain two possible solutions per flux value, with $L_2 > \SI{1}{\micro H}$ being the most consistent with our previous results (see Methods).
Our model remains limited and we do not extract an exact value of $L_2$ from the skewness.
Having a higher $\Icone$ and access to the current-dependent inductance of the JJs could contribute to a more accurate analysis of the magnetic interference patterns.
Finally, from Fig.~\ref{fig:2}i we observe that for a vanishing $L_2$, the expected $\Delta \Phi_{\mathrm{sk}}$ also tends to zero.
Therefore, when measuring interference patterns at critical currents above $\SI{9}{nA}$ no skewness originating from inductance effects is expected.
This is crucial for the next part, where we interpret the current-phase relation measured from an asymmetric configuration of the SQUID.

The shape of the current--phase relation (CPR) in semiconductor weak links is of interest, because it can reveal information about the number of channels through the junction as well as about their transmission~\cite{Beenakker1992}.
In particular, a skewed CPR (having ruled out that the skewness is of inductive origin), is the main indication of a short, highly transmissive, junction~\cite{Della_Rocca2007}.
To probe that, we focus on the asymmetric regime of our device, where we use the flux through the ring to control the superconducting phase difference $\varphi$ across JJ1.
In the most asymmetric regime, we have a critical current ratio of $\Ictwo/\Icone \sim 20$.
All the phase drop imposed by the flux threading the SQUID therefore drops across JJ1~\cite{Della_Rocca2007}, while the moderate $\Ictwo$ is not fully shunting JJ1~\cite{Tinkham2004}.
The phase of JJ1 is then related to the applied flux through $\varphi = 2\pi(\Phi/\Phi_0)$, the CPR being $I(\varphi)$~\cite{Della_Rocca2007}.
Finally, the absence of inductance-induced skewness ensures that the oscillations in critical current correspond to the CPR of JJ1.

\begin{figure}[t]
\centering
\includegraphics[width=0.5\textwidth]{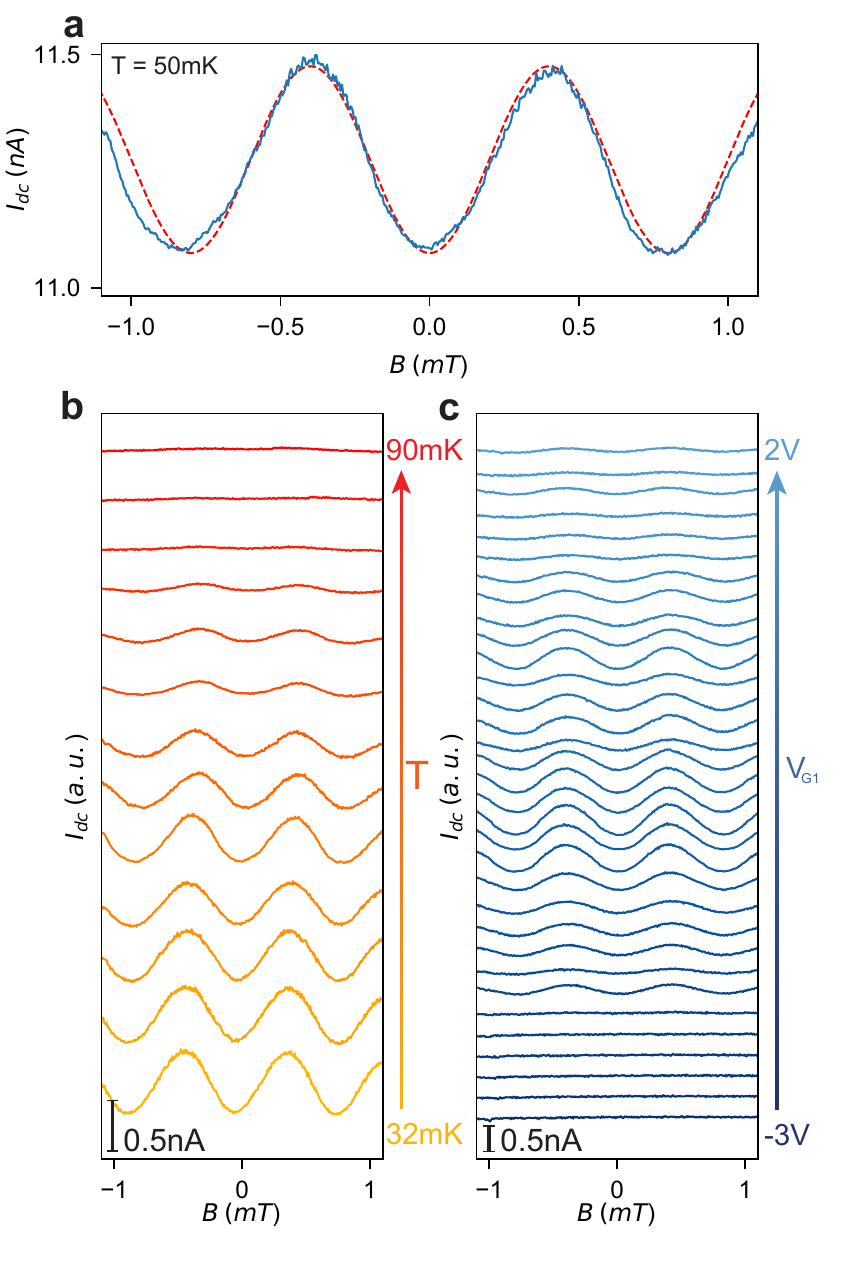}
\caption{\textbf{Measurement of the Current-Phase Relation. a} Current-Phase Relation of JJ1 at base temperature ($T_{\mathrm{base}} = \SI{50}{mK}$) and $\Vgone = \SI{-0.95}{V}$. The CPR is sinusoidal. The red dashed line corresponds to an ideal sine whose periodicity, offset and amplitude have been manually tuned to fit the CPR. \textbf{b} Same CPR taken at $\Vgone = \SI{-0.95}{V}$ at temperatures ranging from $T = \SI{32}{mK}$ to $T = \SI{90}{mK}$ in steps of $\SI{3}{mK}$ for the first one and $\SI{5}{mK}$ for the rest. \textbf{c} Same CPR taken a different values of $\Vgone$ ranging from $\Vgone = \SI{-3}{V}$ to $\Vgone = \SI{2}{V}$ in steps of $\SI{150}{mV}$.}
\label{fig:3}
\end{figure}

We present the CPR measurement result in Fig.~\ref{fig:3}a, taken at the same electrostatic configuration as in Fig.~\ref{fig:1}d, and observe agreement with a sinusoidal modulation (red dashed line), where we attribute the mismatch at higher $B$ field values to the magnetic field dependence of the critical current in each arm.
As shown in Fig.~\ref{fig:3}b, the measured oscillations fade away at $T_{\mathrm{c,JJ1}} = \SI{80}{mK}$.
At our lowest temperature of $\SI{32}{mK}$ (see Methods) no skewness is observed either.
The lack of skewness could be caused by the temperature, the geometry or the diffusivity of JJ1.
If the thermal energy $k_\mathrm{B} T$ is of the order of the proximity-induced gap in the junction region, any skewness in the CPR is expected to fade away~\cite{Beenakker1992}.
Assuming that the temperature is the only reason for which no skewness is observed, the local superconducting gap should be smaller than $\Delta \sim \SI{2.5}{\micro eV}$ to account for the sinusoidal CPR at $\SI{32}{mK}$ (see Supplementary Information).
Such a value is unexpectedly low for the superconducting gap~\cite{Rodan-Legrain2021, Oh2021}, and therefore we dismiss this as a possible explanation.
To investigate the influence of the geometry and diffusivity, we perform CPR measurements at all available $\Vgone$. 
We observe that while the oscillation amplitude is tuned the CPR remains sinusoidal (Fig.~\ref{fig:3}c).
Since we do not expect to change the length of the junction (it is defined by lithography), the change in amplitude could be due to a decrease in transmission as well as a decrease in the number of channels.
The lack of skewness of this CPR excludes JJ1 to be in the short junction limit---the superconducting coherence length is bigger than the junction length---and it having highly transmissive Andreev states at the same time~\cite{Beenakker1992}.
However, having either a long junction or low transmissive channels (or both), results in a sinusoidal CPR as observed. 
Based on our observation we cannot rule out any of these scenarios and therefore hope for future experiments on shorter and more transmissive junctions. 

In conclusion, we have shown a monolithic superconducting quantum interference device in MATBG, allowing for local and all-electrical tuning of its critical current and inductance.
Our SQUID exhibits phase coherence over long distances and testifies to the realization of a superconducting-phase-controlled experiment in this material.
The device geometry has allowed us to extract the effective charge $2e$ for the superconducting charge carriers, to observe a trend in the superfluid density with temperature, and to directly measure the current-phase relation of one of its JJs.
By measuring a kinetic inductance of up to $\SI{2}{\micro H}$ at base temperature, we have shown the potential of the material to host devices requiring high inductances such as phase-slip junctions or high-inductance photon detectors.
The combination of local, all-electrical tunability and phase control make our device a probe of MATBG complementary to previously reported techniques and a stepping stone for complex \textit{in situ}-tunable superconducting architectures.

\section{Data availability}

The data that support the findings of this study will be made available online through the ETH Research Collection.

\section{Acknowledgements}
We thank Peter M\"{a}rki, Lev Ginzburg, Petar Tomić and the staff of the ETH cleanroom facility FIRST for technical support. We thank Hugues Pothier for helpful and detailed discussions on superconducting devices and Oded Zilberberg, Jared Cole and members of the quantum e-leaps consortium for comments on our data.
We acknowledge financial support by the European Graphene Flagship, the ERC Synergy Grant Quantropy, the European Union's Horizon 2020 research and innovation program under grant agreement number 862660/QUANTUM E LEAPS and NCCR QSIT (Swiss National Science Foundation).
K.W. and T.T. acknowledge support from the Elemental Strategy Initiative conducted by the MEXT, Japan (Grant Number JPMXP0112101001) and  JSPS KAKENHI (Grant Numbers 19H05790, 20H00354 and 21H05233).
E.P. acknowledges support of a fellowship from ”la Caixa” Foundation (ID 100010434) under fellowship code LCF/BQ/EU19/11710062.

\section{Author information}
\subsection{Corresponding authors}
Correspondence and requests for materials should be addressed to E.P. or F.K.d.V.

\subsection{Author contributions}
E.P. fabricated the device.
T.T. and K.W. supplied the hBN crystals.
E.P. and F.K.d.V. performed the measurements and analysis of the data.
E.P., F.K.d.V., S.I., G.Z. and P.R. discussed the data.
F.K.d.V., T.I. and K.E. supervised the project.
E.P. and F.K.d.V. wrote the manuscript with comments from all authors.

\subsection{Competing interests}
The authors declare no competing financial interests.

\section{Methods}

\subsection{Twist angle estimation}

For extracting the twist angle of the device, we first relate the electron density at which we observe the full filling of the flat bands to the twist angle of the sample. This relation is the following~\cite{Cao2018_1}:

\begin{equation}
    \theta = 2 \arcsin \left( \frac{a}{2L} \right)
\end{equation}

with L being the distance between two closest AA stacked regions,

\begin{equation}
    L = 2 \sqrt{2 \mathcal{A}/\sqrt{3}} 
\end{equation}

and $\mathcal{A}$ being the area of a moiré unit cell. Each moiré unit cell can host four electrons because of spin and valley degeneracy (we write $g = 4$ for the degeneracy). It is thus related to the electron density in the sample by the following equation:

\begin{equation}
    \mathcal{A} = \frac{g}{n_{\mathrm{filling}}}
\end{equation}

We then read the values for $n_{\mathrm{filling}}$ under each local top gate from the density maps (Extended Data Fig.~\ref{fig:S_D}) and an overall value for the whole sample. This results in approximate twist angles of $0.95^\circ$ for the whole sample, $0.98^\circ$ for the region under gate G1 and $1.02^\circ$ for the region under gate G2.

\subsection{Device fabrication}

We first assemble the so-called stack using the dry pick-up method~\cite{Kim2016}. All the flakes are exfoliated on a $\SI{285}{nm}$ $\mathrm{p:Si/SiO_2}$ wafer. To begin, we cut a large area ($>\SI{40}{\micro m} \times \SI{40}{\micro m}$) graphene flake in two using a tungsten needle of a tip diameter of $\SI{2}{\micro m}$, controlled by a micromanipulator. For the pick up phase we use a self made polydimethylsiloxane/polycarbonate stamp. The top hexagonal boron nitride (hBN) flake, of a thickness of $\SI{25}{nm}$, is picked up first at a temperature of $\SI{80}{C^\circ}$. We then pick up the first half of the previously cut graphene flake at $\SI{40}{C^\circ}$, rotate the stage by $1.1^\circ$ and pick up the other half. Then we move on to picking up the bottom hBN flake, of a thickness of $\SI{85}{nm}$. We first contact it at $\SI{40}{C^\circ}$ and then raise the temperature of the stage to $\SI{80}{C^\circ}$. Finally, we pick up a graphite flake, which will be used as a back gate, by contacting it at $\SI{100}{C^\circ}$. The stack is then deposited at $\SI{180}{C^\circ}$ on a $\mathrm{p:Si/SiO_2}$ chip. We clean the polycarbonate present on the chip and stack after deposition using dichloromethane. We define the area where the edge contacts will be evaporated by electron beam lithography and etch the hBN by reactive ion etching ($\mathrm{CHF_3/O_2}$, 40/4 sccm, 60W, with a hBN etching rate of $\SI{0.6}{nm/s}$). Subsequently, we evaporate the contacts (Cr/Au, 10/70nm). We define, again using electron beam lithography, the lines to the etched contacts (Cr/Au, 10/110nm). The whole stack is etched to define the mesa and deposit a $\SI{30}{nm}$ thick layer of aluminum oxide by atomic layer deposition. We follow this by defining the top gates in our last electron beam lithography step and evaporate them (Cr/Au, 10/110nm).

\subsection{Measurement setup}

In a two-terminal setup, we apply a current bias and measure the corresponding voltage drop across the sample. To generate the bias current, we use an in-house built d.c. source in series with a $\SI{100}{M\Omega}$ resistor. We use a d.c. amplifier built in-house and measure its output with a Hewlett Packard 3441A multimeter. Each gate is connected to a different voltage source, of the same type as the one used for generating a direct current. We convert the voltages we apply to our gates to electron density by a parallel plate capacitor model. We estimate the capacitance per unit area of each gate $i$ to be:

\begin{equation}
    C_i = \sum_{j} \varepsilon_0 \times \varepsilon_j / d_j
\end{equation}

where $\varepsilon_0$ is the vacuum permittivity, and $\varepsilon_j$ and $d_j$ are, respectively, the relative permittivity and thickness of layer $j$. We then calculate the electron density $n$ as follows:

\begin{equation}
    n = \sum_{i} C_i V_i / e,
\end{equation}

$e$ being the elementary charge and $V_i$ the voltage applied at gate $i$. We verify that the capacitance per unit area that we estimate for the back gate fits the Landau Fan dependence on density (Extended Data Fig.~\ref{fig:S_G})
\\

In order to extract the critical current as a function of magnetic field, we continuously sweep the bias current and readout the voltage increase corresponding to the switching out of the superconducting state only, instead of measuring the voltage drop at every point in current bias. This is done with a device built in-house which consists of an analog circuit in combination with a digital pyboard controller.
Before performing such a measurement, we measure (with the 'regular' measurement setup) an I/V curve and readout the jump in voltage when switching out of the superconducting state. We then set the threshold $V_{\mathrm{th}}$ to be the average value of the voltages defining such voltage jump. For measuring, we linearly ramp up the current across the device, starting at a negative value, resulting in a negative voltage being readout at the beginning of the measurement procedure. As the current is ramped up towards positive values, the voltage drop changes sign. This sign change triggers the pyboard, which stores the time at which it took place. As the current keeps rising up linearly, so does the voltage drop, which ends up attaining the threshold value $V_{\mathrm{th}}$, stored in the pyboard memory. The controller is then triggered again, and stores the time at which the event took place and permutes a series of digitally controlled switches in the analog circuit of the device. This leads to a fast ramp down of the current to the negative value at the beginning of the measurement procedure. Then, from knowing the ramp-up speed of the current and the time difference between the two triggers, we deduce the current at which the superconducting transition took place. It is important to note that this measurement can only be performed in regimes where the switching out transition presents a sharp step in voltage. This ensures that there is a wide enough range of values for the voltage threshold that we choose which lead to the same resulting critical current. We are therefore certain that we do not introduce modifications to physically meaningful data by our choice of a voltage threshold.

All measurements are performed in a $^3$He--$^4$He dilution refrigerator with a base temperature of $\SI{50}{mK}$. To temporarily reach lower temperatures, we perform a so-called single-shot procedure in which the flow of $^3$He returning from the mixing chamber (MC) is diverted to the dump of the cryostat instead of the condenser line. This interrupts the flow of liquid (and warmer) $^3$He from the condenser towards the MC, preventing if from exchanging heat with the colder outgoing flow of $^3$He. The procedure results in a lower heat load for the still and MC, leading to a decrease in temperature of the MC. The state can only hold as long as there is $^3$He in the MC, of which the incoming flow has been interrupted. Once all the $^3$He is removed from the MC, the system warms up.

\subsection{Offset in magnetic field}
We observe an offset in magnetic field due to our magnet. In addition, as shown in Extended Data Fig.~\ref{fig:S_H}, the offset can depend on the history of the applied field. For consistency through the manuscript, we subtract for each measurement an offset so that the magnetic interference pattern results in a symmetric plot. Because of such determination of the zero field, we do not make any conclusion from the value of the critical current at zero field.

\subsection{Estimation of the geometric inductance}

To estimate the geometric contribution to the total inductance of the device we model our SQUID as a rectangular loop of wire. The formula giving the inductance is then~\cite{Grover1962}:

\begin{eqnarray*}
    L_{\mathrm{geo}} = \frac{\mu_0 \mu_r}{\pi}
    \big[-2(w+h)+2\sqrt{h^2+W^2} \\
    -h \ln \left( \frac{h + \sqrt{h^2+W^2}}{W} \right)\\
    -W\ln\left( \frac{W + \sqrt{h^2+W^2}}{h} \right) \\
    +h\ln\left( \frac{2h}{a}\right) \\
    +W\ln\left( \frac{2W}{a}\right) \big]
\end{eqnarray*}

with $W$ being the length of the loop, $h$ its height, $a$ its radius and $\mu_\mathrm{r}$ the relative permitivity of the material. We set $W = \SI{4}{\micro m}$, $h = \SI{2}{\micro m}$, $a = \SI{1}{nm}$ and $\mu_\mathrm{r} = 1$ and obtain $L_{\mathrm{geo}} = \SI{17}{pH}$, which is many orders of magnitude away from the values measured. Having set a bigger radius for our wire in this calculation would have resulted in an even smaller value. Therefore, we attribute the inductance of the device mainly to its kinetic inductance.

\subsection{Relation of the skewness of the magnetic interference pattern to the macroscopic parameters of the SQUID}

When operating a SQUID close to the critical current symmetric regime, the eventual skewness of a magnetic interference pattern is related to the inductance asymmetry $\alpha_\mathrm{L} = (L_1 - L_2)/(L_1 + L_2)$ through the following relation \cite{Tesche1977}:

\begin{equation}
    \Delta \Phi_{\mathrm{sk}} = \beta \alpha_\mathrm{L} i_{\mathrm{sk}},
\end{equation}

with $\Delta \Phi_{\mathrm{sk}}$ the flux shift at the maximal critical current, in units of the flux quantum, $\beta = (\Icone + \Ictwo)L/\Phi_0$ and $i_{\mathrm{sk}}$ the current at which $\Delta \Phi_{\mathrm{sk}}$ is measured in units of the average critical current of the arms, here leading to $i_{\mathrm{sk}} = 2$. Having $L = L_1 L_2 / (L_1 + L_2)$, we fix $L_1 = \SI{2}{\micro H}$ and plot $\Delta \Phi_{\mathrm{sk}}$ as a function of $L_2$ in Fig.~\ref{fig:2}i.
Two values of $L_2$ are possible for a range of $\Delta \Phi_{\mathrm{sk}}$.
After estimating $\Delta \Phi_{\mathrm{sk}}$ of the order of a few percent we choose $L_2 > \SI{1}{\micro H}$ by exclusion: A value of $L_2$ an order of magnitude lower than $L_1$ in the symmetric current regime would result in $\Delta \Phi \sim \Phi_0$ according to Eq.~\eqref{eq:inductance}, which we do not observe in Fig.~\ref{fig:2}d.
The value $L_2 > \SI{1}{\micro H}$ is higher than those obtained for critical currents higher than $\SI{3}{nA}$. The inductance $L_2$ is not accessible through Eq.~\eqref{eq:single_inductance} at low values of $\Ictwo$ due to lack of resolution of our data. 
However, following $L_\mathrm{K} \propto m^*/(n_\mathrm{s}\mathcal{A})$, one could expect the inductance to increase non-linearly as the cross-section of the supercurrent tends to zero, which happens when locally tuning the material under top gate 1 to the edge of the superconducting regime.

\clearpage
\onecolumngrid

\begin{center}
    \Large{\textbf{Supplementary Information}}
\end{center}

\section{Heating effect in the critical current oscillations}

As shown in Extended Data Fig.~\ref{fig:S_A}, for the most asymmetric regime, we observe oscillations in the switching current as a function of magnetic field, but not in the retrapping current. We attribute this to a heating effect. When the SQUID is being biased above the critical current, a voltage drop develops across each arm. This results in an electron temperature being higher than the temperature of the cryostat~\cite{Courtois2008}. When the retrapping into the superconducting state takes place, the electron temperature in arm 1 is too high to allow for a transition into the superconducting state, preventing the formation of a ring until superconductivity has been developed in arm 2. The absence of a ring during the transition prevents phase coherence around the etched hole, thus preventing oscillations in the critical current from appearing. When the critical current asymmetry is lower, the voltage drop across arm 1 when the transition takes place is lower, thus the electron temperature is also lower. Gradually, as symmetry in $I_\mathrm{c}$ is restored, this reduction in electron temperature leads to an appearance of oscillations also in the retrapping current.

\clearpage
\newpage

\section{Premature switching of a JJ due to thermal effects}

We neglect all the inductive effects in this example and call $\delta_1$ the phase across JJ1, $\delta_2$ the phase across JJ2 and $\phi = 2\pi(\Phi/\Phi_0)$ the phase around the loop due to the applied magnetic flux. We thus have $\phi = \delta_2 - \delta_1$.

From the measured CPR in our device, we assume it to be sinusoidal in this example, yielding a total current through the device $I = \Icone \sin(\delta_1) + \Icone \sin(\delta_2)$. We now call $\delta_{\mathrm{2,sw}}$ the phase of JJ2 when the switching to the normal state actually occurs in arm 2. Then, the positive switching current is 
\begin{equation*}
    I_{+} = \Icone \sin(\delta_{\mathrm{2,sw}} - \phi) + \Ictwo \sin(\delta_{\mathrm{2,sw}})
\end{equation*}
and the negative one is
\begin{eqnarray*}
    I_{-} = \Icone \sin(-\delta_{\mathrm{2,sw}} - \phi ) + \Ictwo \sin(-\delta_{\mathrm{2,sw}}) \\
    = -\Icone \sin(\delta_{\mathrm{2,sw}} + \phi) - \Ictwo \sin(\delta_{\mathrm{2,sw}}).
\end{eqnarray*} 

Since $\Ictwo >> \Icone$ the maximum of $I_{+}$ is reached when JJ2 switches out of the superconducting state, and $\delta_{\mathrm{2,sw}} - \phi_+ = \pi/2$ thus at $\phi_+ = \delta_{\mathrm{2,sw}} - \pi/2$. On the other hand, the maximum in $|I_{-}|$ is reached when $\delta_{\mathrm{2,sw}} + \phi_- = \pi/2$ thus at $\phi_- = \pi/2 - \delta_{\mathrm{2,sw}}$. Therefore, the offset in phase due to JJ2 switching before reaching a phase of $\pi/2$ is
\begin{equation*}
    \Delta \phi_{\mathrm{sw}} = \phi_+ - \phi_- = 2\delta_{\mathrm{2,sw}} - \pi.
\end{equation*}
In the extreme case in which JJ2 switches at $\delta_{\mathrm{2,sw}} = 0$, we have $\Delta \phi_{\mathrm{sw}} = -\pi$. The measured offset in magnetic field reaches $\Delta \Phi \sim \Phi_0$, equivalent to $\Delta \phi \sim 2\pi$ (Fig.~\ref{fig:2}b). Therefore, in the worst case scenario, at least half of this offset comes from the inductance of the loop.

This premature switching effect can only be measured if being able to determine the critical current of the junction, by which we mean its switching current at zero temperature~\cite{Della_Rocca2007}. Unfortunately, such analysis only holds for superconductor-insulator-superconductor (SIS) JJs~\cite{Devoret1985} and cannot be performed in our superconductor-normal-superconductor (SNS) JJs. However, when performing repetition measurements of the switching current of the JJ of a device hosting a similar type of JJ~\cite{deVries2021}, we observe a small broadening of the current histogram due to thermal effects, comparing it to the ones typically obtained in the SIS case. We therefore neglect the contribution of this effect to the measured offset in magnetic field when performing our analysis on the inductances of the device.

\clearpage
\newpage

\section{Estimation of the induced superconducting gap from the current-phase relation}

The CPR of a JJ can be derived from its free energy~\cite{Beenakker1992} $F$:

\begin{equation}
    I(\varphi) = \frac{2e}{h} \frac{\partial F}{\partial \varphi},
\end{equation}

where $e$ is the elementary charge, $\varphi$ the phase drop across the junction and $h$ Planck's constant. In the so-called short-junction limit and considering the superconducting gap $\Delta$ to be constant this leads to the following expression for the CPR at a temperature $T$:

\begin{equation}
    I(\varphi) = \frac{e \Delta ^2 \sin(\varphi)}{2 \hbar} \sum_{p} \frac{\tau_p}{\varepsilon_p(\varphi)} \tanh\left(\frac{\varepsilon_p(\varphi)}{2 k_\mathrm{B}T} \right),
 \end{equation}
 
where $\hbar$ is the reduced Planck's constant and $k_\mathrm{B}$, the Boltzmann constant, $\tau_p$ the transmission of the $\mathrm{p^{th}}$ channel and $\varepsilon_p(\varphi) = \Delta \sqrt{1 - \tau_p \sin^2(\varphi/2)}$ the phase-dependent energy of Andreev states.
From this formula, we observe that if the hyperbolic tangent function is small enough to be in the range where $\tanh(x) \sim x$, the argument of the sum over Andreev states is 1, making of the current a \textit{non-skewed} sum of sinus terms.
Assuming that the non-skewness of the CPR is due to temperature effects therefore corresponds to having $\tanh(x) \sim x$.
For our estimations, we set $\varepsilon_p(\varphi) = \Delta$ and the resulting argument of the hyperbolic tangent $\Delta/2 k_\mathrm{B}T$ to $1/2$ ($\tanh{0.5} = 0.46$). Which leads to $\Delta = k_\mathrm{B}T$, thus obtaining $\Delta \sim \SI{2.5}{\micro eV}$ for $T = \SI{30}{mK}$. As mentioned in the main text, we dismiss such an unexpectedly low value as the cause for the sinusoidal shape of the measured CPR.

\clearpage

\begin{center}
    \Large{\textbf{Extended data}}
\end{center}
\setcounter{figure}{0}
\renewcommand{\thefigure}{S\arabic{figure}}
\renewcommand{\theequation}{S\arabic{equation}}
\setcounter{secnumdepth}{2}

\begin{figure*}[!h]
\centering
\includegraphics[width=0.5\textwidth]{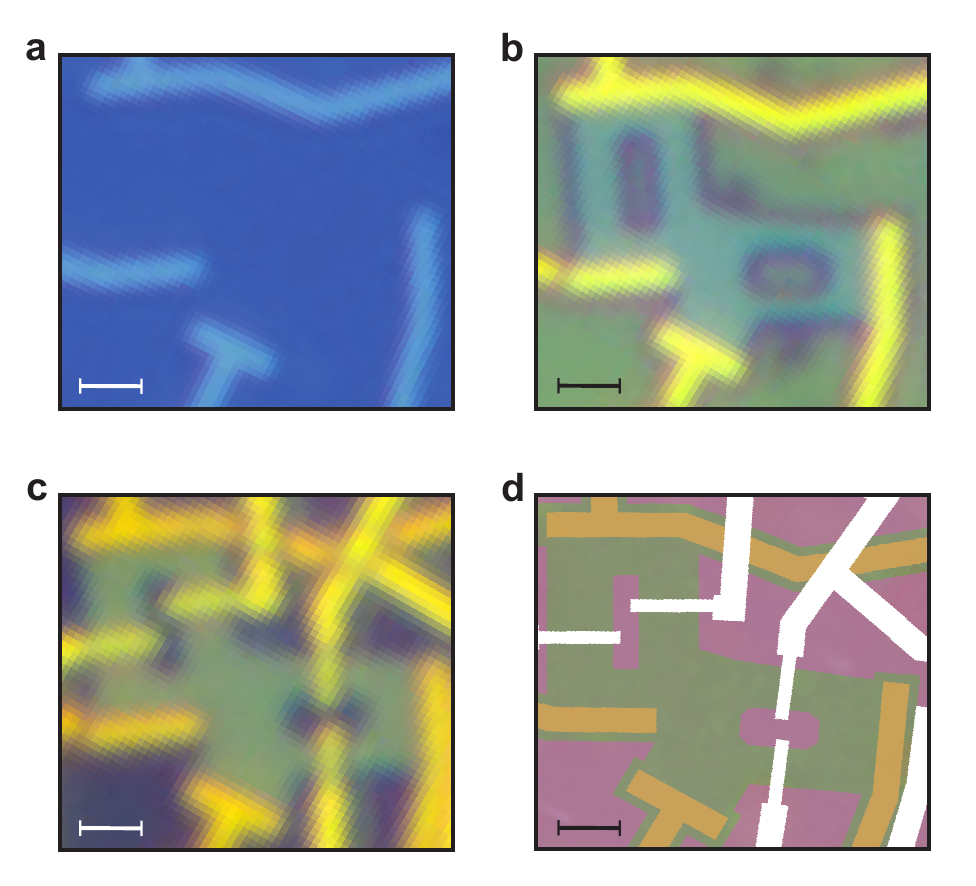}
\caption{\textbf{Device fabrication: Capture of the design file and optical images taken during the fabrication process.} \textbf{a} Optical image of the device after etching and evaporation of the gold contacts. The scale bar corresponds to a length of $\SI{1}{\micro m}$. \textbf{b} Optical image of the device after mesa etching. \textbf{c} Optical image of the device after the deposition of a layer of aluminum oxide and definition and evaporation of top gates. \textbf{d} Capture of the design file of the device. Contacts are depicted in yellow, etched areas in pink and top gates in white.}
\label{fig:S_C}
\end{figure*}

\clearpage
\newpage

\begin{figure*}[!h]
\centering
\includegraphics[width=1\textwidth]{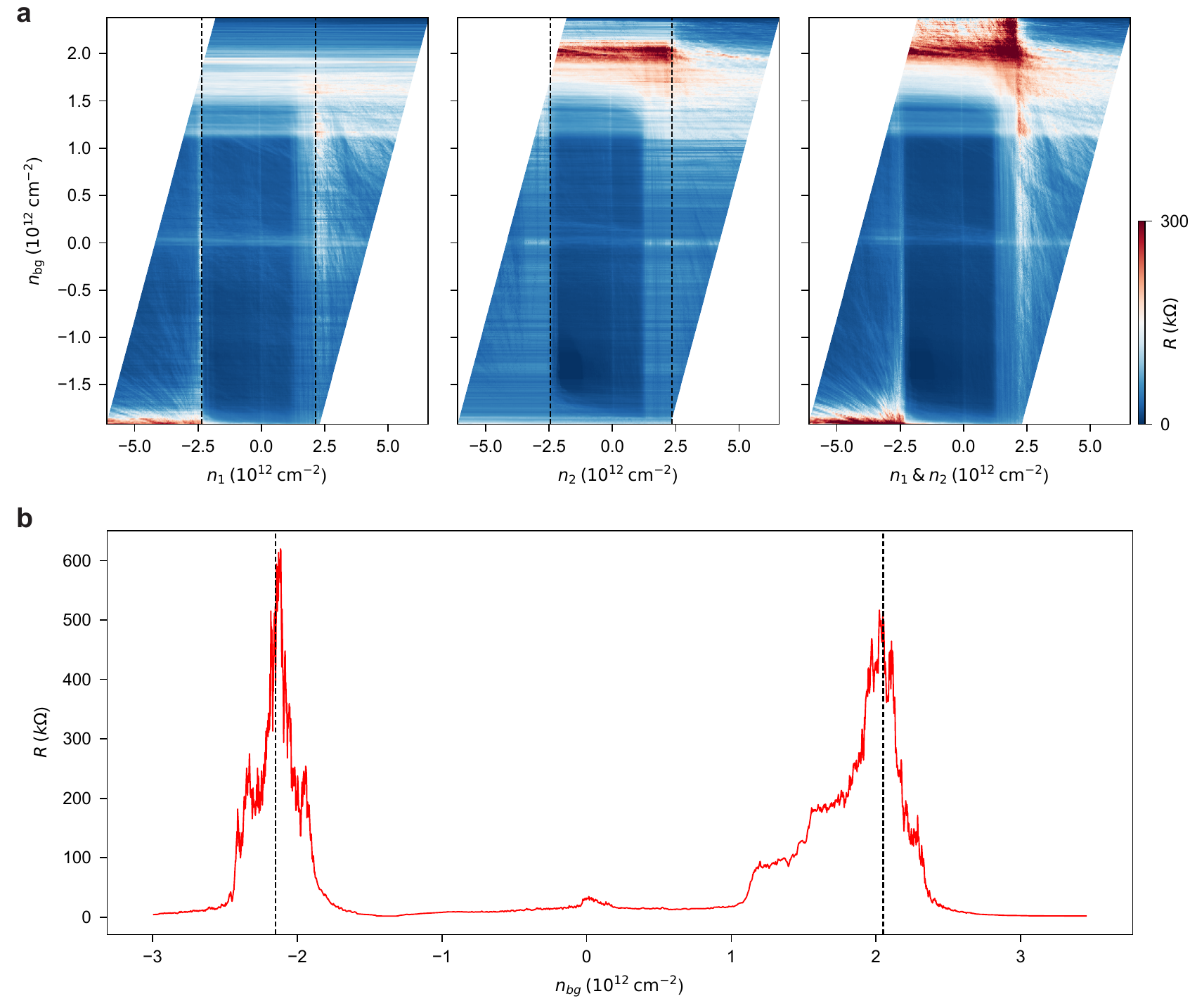}
\caption{\textbf{Density maps and angle extraction. a} Differential resistance maps across the device as a function of electron density in regions biased by only the back gate (global) or both back and top gates. The data in the left panel corresponds to $\Vgone$ being swept and gate 2 being fixed at $\Vgtwo = \SI{10}{V}$. In the center panel $\Vgtwo$ is swept while gate 1 is kept constant at $\Vgone = \SI{10}{V}$. In the right panel both top gates are swept together. Densities are computed from the model described in the Methods section. Black dashed lines indicate the values at which we consider the flat bands to be fully filled. These values are then used to extract the twist angle of the device. \textbf{b} Extended range trace of the resistance as a function of the global density induced by the back gate, with each local top gate set to zero.}
\label{fig:S_D}
\end{figure*}

\clearpage
\newpage

\begin{figure*}[!h]
\centering
\includegraphics[width=0.8\textwidth]{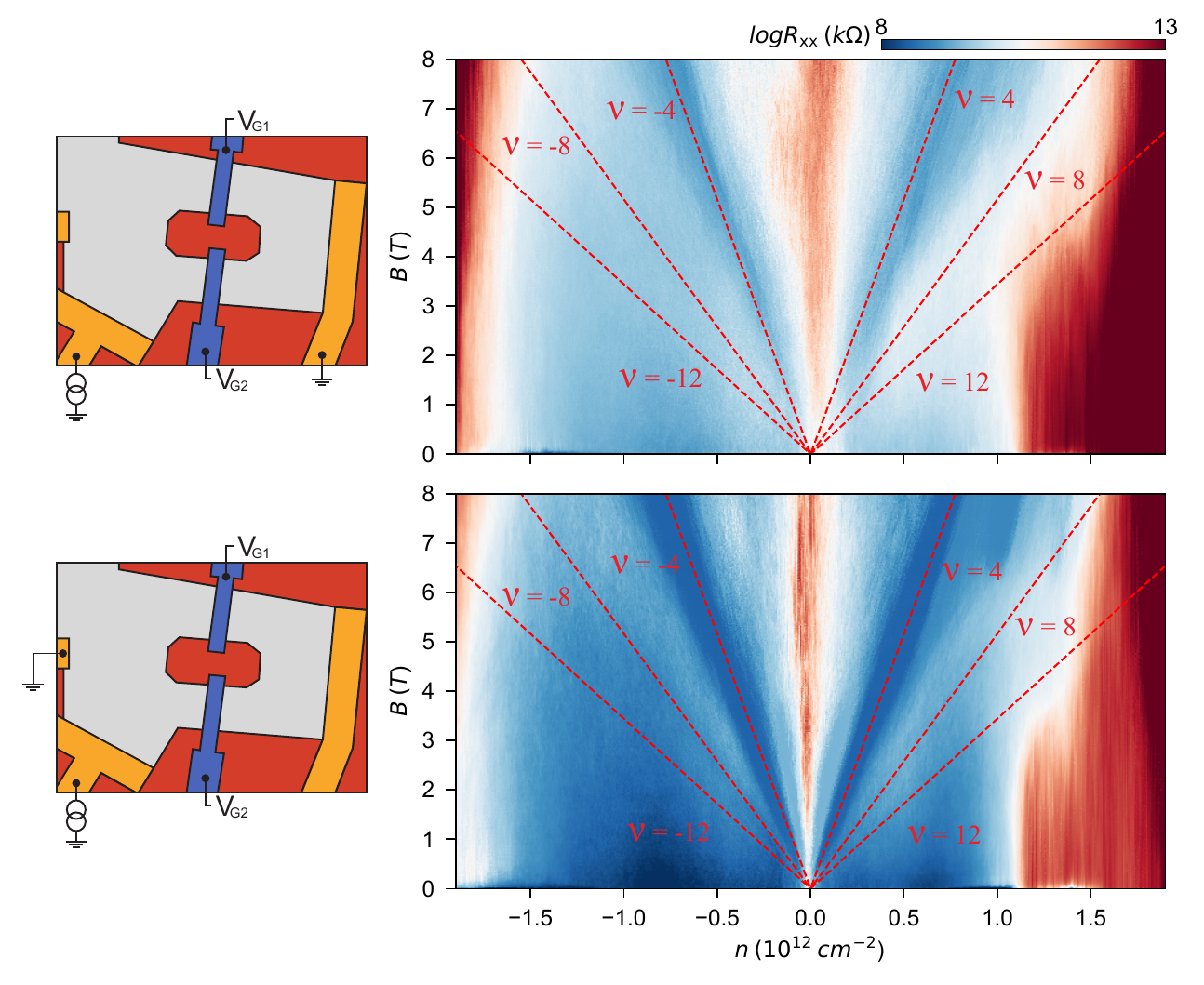}
\caption{\textbf{Landau Fans at different contact configurations.} The top panel corresponds to a Landau Fan across the device using the same contact configuration as used in all the measurements shown in the main text. The bias current is $I_{\mathrm{dc}} = \SI{1}{nA}$. Red dashed lines correspond to the expected Landau levels (their corresponding filling factors are indicated) from our estimation of capacitance per unit area between the back gate and the graphene. The bottom panel shows the same measurement taken between the two contacts that are not separated by the etched hole.}
\label{fig:S_G}
\end{figure*}

\clearpage
\newpage

\begin{figure*}[!h]
\centering
\includegraphics[width=0.8\textwidth]{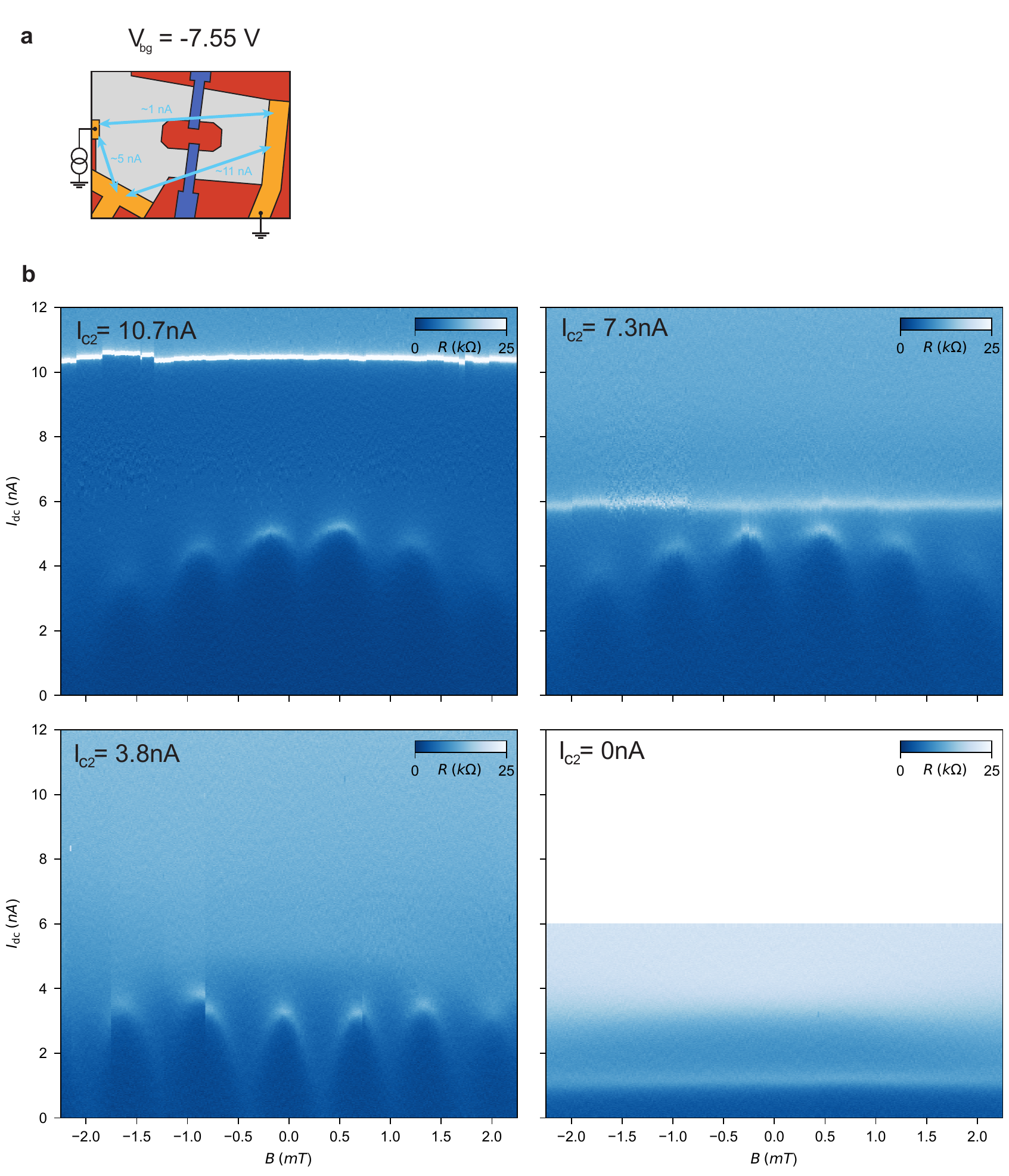}
\caption{\textbf{Magnetic interference between contacts I and III.} \textbf{a} Schematics of the device, with the approximate maximal critical current that can be reached at $V_{\mathrm{bg}} = \SI{-7.55}{V}$ between each pair of contacts. \textbf{b} Differential resistance between contacts I and III as a function of current bias and magnetic field. The gate configuration is the same as for Fig.~\ref{fig:1}d. $\Vgtwo$ is then tuned to reduce the critical current of arm 2, as indicated in the top right corner of each panel of the figure. We observe that at the maximal critical current, a transition at $\SI{10.7}{nA}$ without oscillations is present, as well as an interference pattern at lower current values. As we decrease $\Ictwo$ to $\SI{7.3}{nA}$ the transition at higher currents goes to lower currents with the interference pattern unchanged. When $\Ictwo$ goes below the current at which the maximum of the interference pattern is observed, we see a modification of such pattern. Finally, when superconductivity is switched off in arm 2, we observe no superconducting pattern. We deduce from this that, at the optimal back gate voltage for contact combination II-III (main text) there is an area of arm 2 in the configuration I-III with $I_\mathrm{c} \sim \SI{5}{nA}$, limiting to this value the maximal current attainable by the interference pattern.}
\label{fig:S_I}
\end{figure*}

\clearpage
\newpage

\begin{figure*}[!h]
\centering
\includegraphics[width=0.8\textwidth]{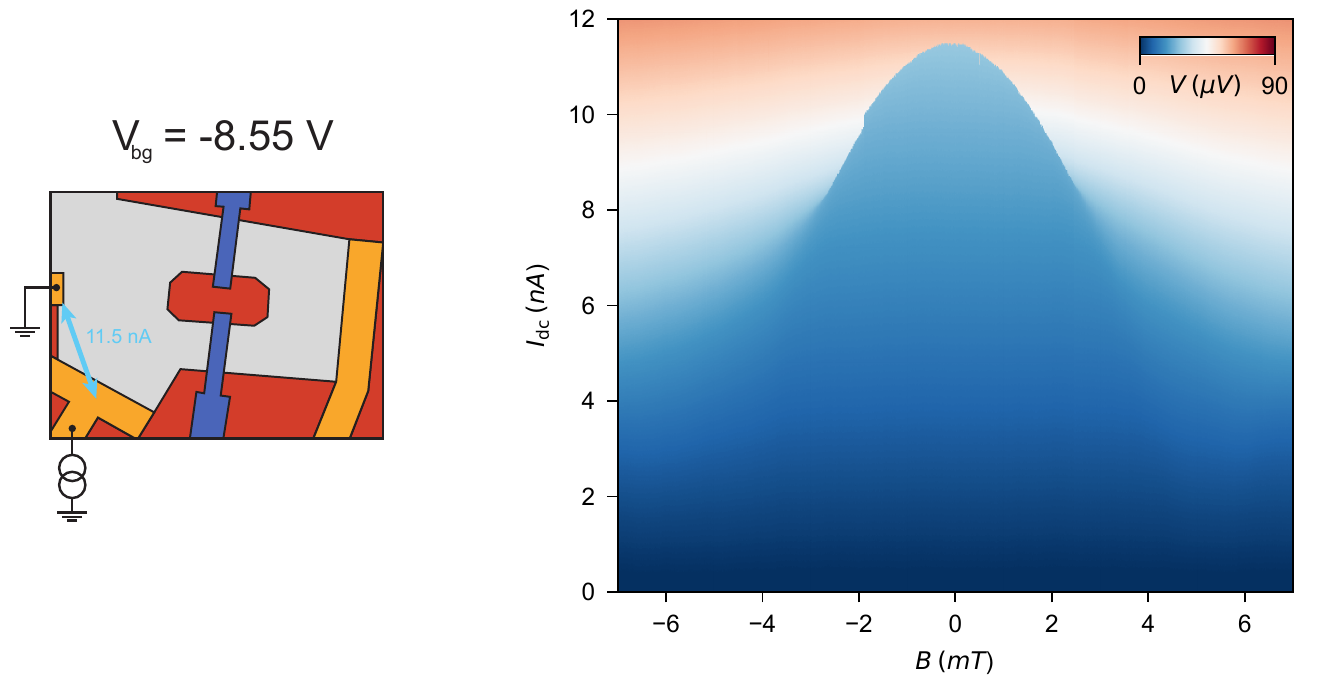}
\caption{\textbf{Critical current as a function of magnetic field between contacts I and II.} \textbf{a} Schematics of the device, with the approximate maximal critical current that can be reached at $V_{\mathrm{bg}} = \SI{-8.55}{V}$ between the pair of contacts. \textbf{b} Voltage drop between contacts I and II (no hole between them) as a function of current bias and magnetic field. The transition region is the critical current. We observe no oscillations in critical current, further confirming the interference in data shown in the main text and ruling out the effect of a second etched loop (see Extended Data Fig.~\ref{fig:S_C}.)}
\label{fig:S_J}
\end{figure*}

\clearpage
\newpage

\begin{figure*}[!h]
\centering
\includegraphics[width=0.5\textwidth]{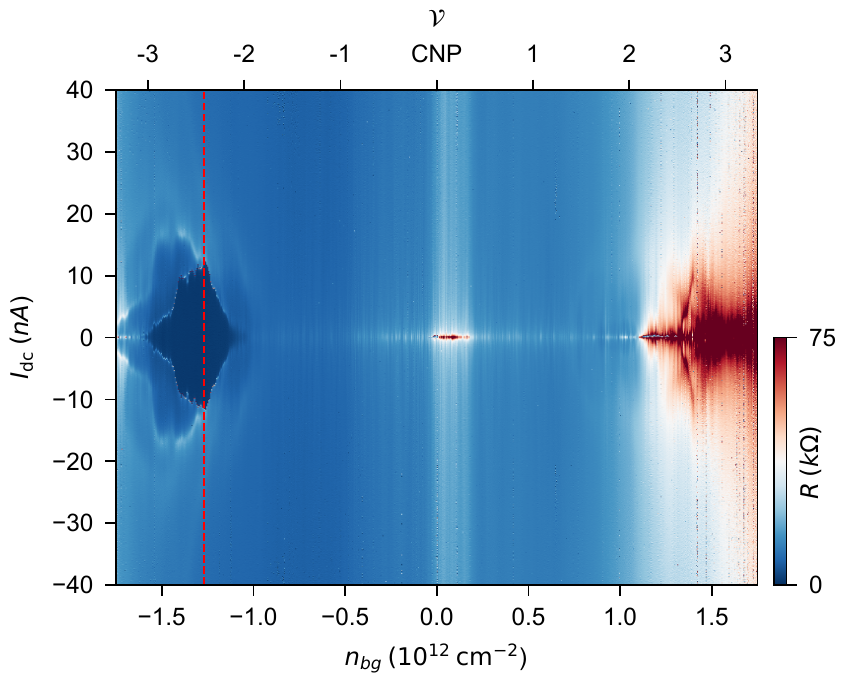}
\caption{\textbf{Resistance of the device as a function of density.} $\Vgone$ and $\Vgtwo$ are kept at zero volts. We plot the derivative of the measured voltage drop as a function of the current bias, which corresponds to the differential resistance $R=dV/dI_\mathrm{dc}$ across the device. The red dashed line indicates the 'optimal density' $n_\mathrm{opt} = \SI{-1.27e12}{cm^{-2}}$ at which we set the back gate when wanting to maximize the critical current through the device. The upper axis indicates the filling factors corresponding to the electron densities.}
\label{fig:S_E}
\end{figure*}

\clearpage
\newpage

\begin{figure*}[!h]
\centering
\includegraphics[width=0.75\textwidth]{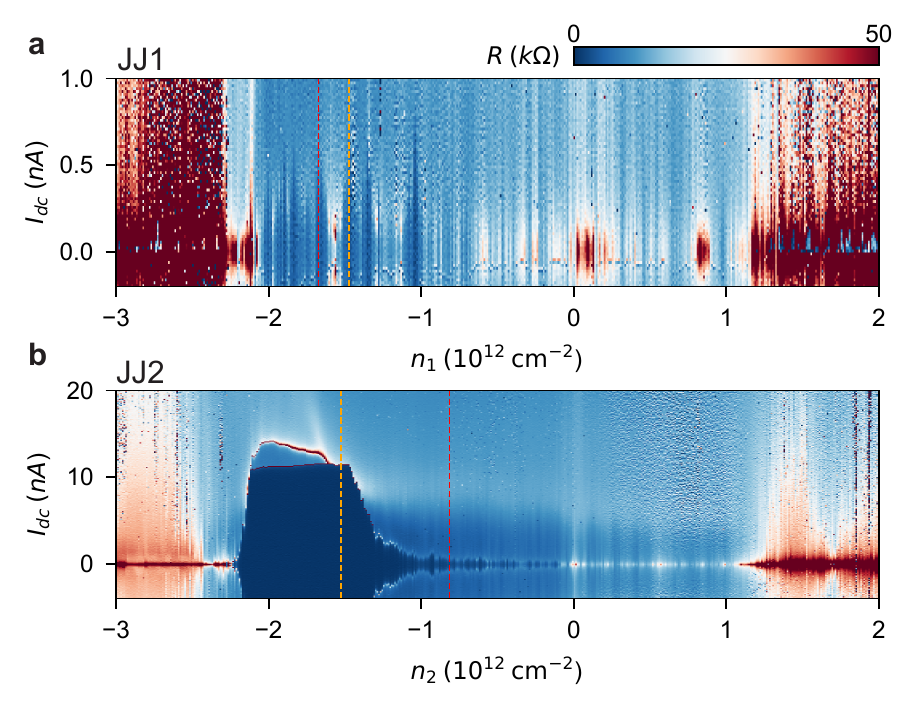}
\caption{\textbf{Differential resistance across each arm of the loop as a function of local densities.} The back gate is tuned to its optimal point $n_\mathrm{opt} = \SI{-1.27e12}{cm^{-2}}$. We step the local top gate of each arm and sweep the current bias. For \textbf{a}, the voltage bias of gate G1 is fixed at $V_{\mathrm{G2}} = \SI{10}{V}$, to prevent any supercurrent from going through it, while we step $V_{\mathrm{G1}}$. For \textbf{b} the configuration is the opposite, we have $V_{\mathrm{G1}} = \SI{10}{V}$ while we step $V_{\mathrm{G2}}$. The dashed lines correspond to the voltage of each top gate at the asymmetric (orange) and symmetric (red) regime measurements shown in Fig.~\ref{fig:1}d,e.}
\label{fig:S_F}
\end{figure*}

\clearpage
\newpage

\begin{figure*}[!h]
\centering
\includegraphics[width=1\textwidth]{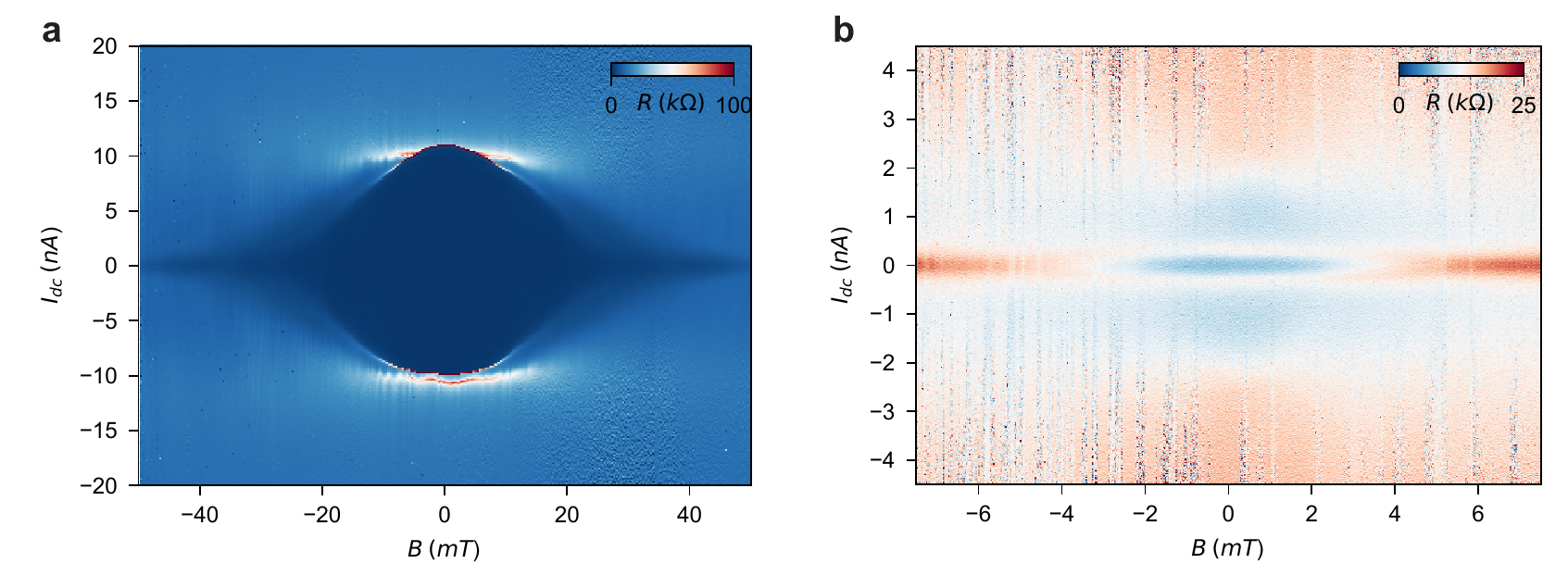}
\caption{\textbf{Critical field at each arm of the device.} The voltage of the back gate is set to the optimal point for every figure. \textbf{a} Differential resistance as a function of current bias and magnetic field when $\Vgone = \SI{10}{V}$, preventing any supercurrent from flowing through arm 1 and $\Vgtwo = \SI{-0.5}{V}$ maximizes the critical current of arm 2. We thus observe the resistance across arm 2 of the device as a function of current and magnetic field. \textbf{b} Same measurement as in \textbf{a}, this time with $\Vgtwo = \SI{10}{V}$ and $\Vgone = \SI{-0.37}{V}$.}
\label{fig:S_B}
\end{figure*}

\clearpage
\newpage

\begin{figure*}[!h]
\centering
\includegraphics[width=1\textwidth]{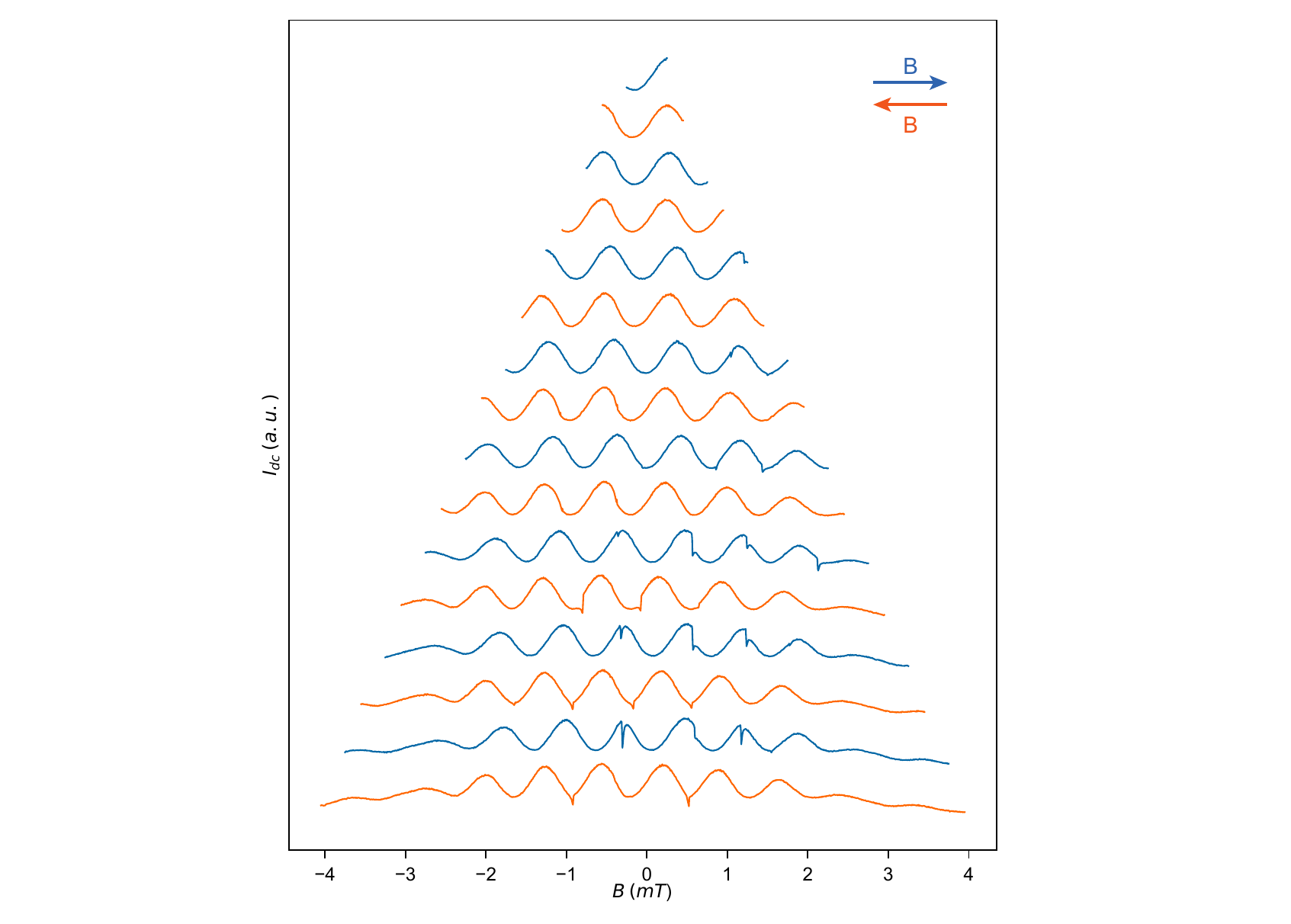}
\caption{\textbf{Hysteresis and discontinuities in the critical current as a function of magnetic field.} Each line represents a CPR trace taken at the most asymmetric regime of the device. The magnetic field is swept from negative to positive in a small range, then the direction is reversed and the range increased. This procedure is repeated several times to obtain the data shown in the figure. Switches in the CPR traces and hysteresis appear as the range of the magnetic field sweep increases. We can not rule out a ferromagnetic part of the cryostat or the superconducting magnet to be at the origin of this phenomenon. Therefore, we do not highlight it in the main text.}
\label{fig:S_H}
\end{figure*}


\clearpage
\newpage

\begin{figure*}[!h]
\centering
\includegraphics[width=0.8\textwidth]{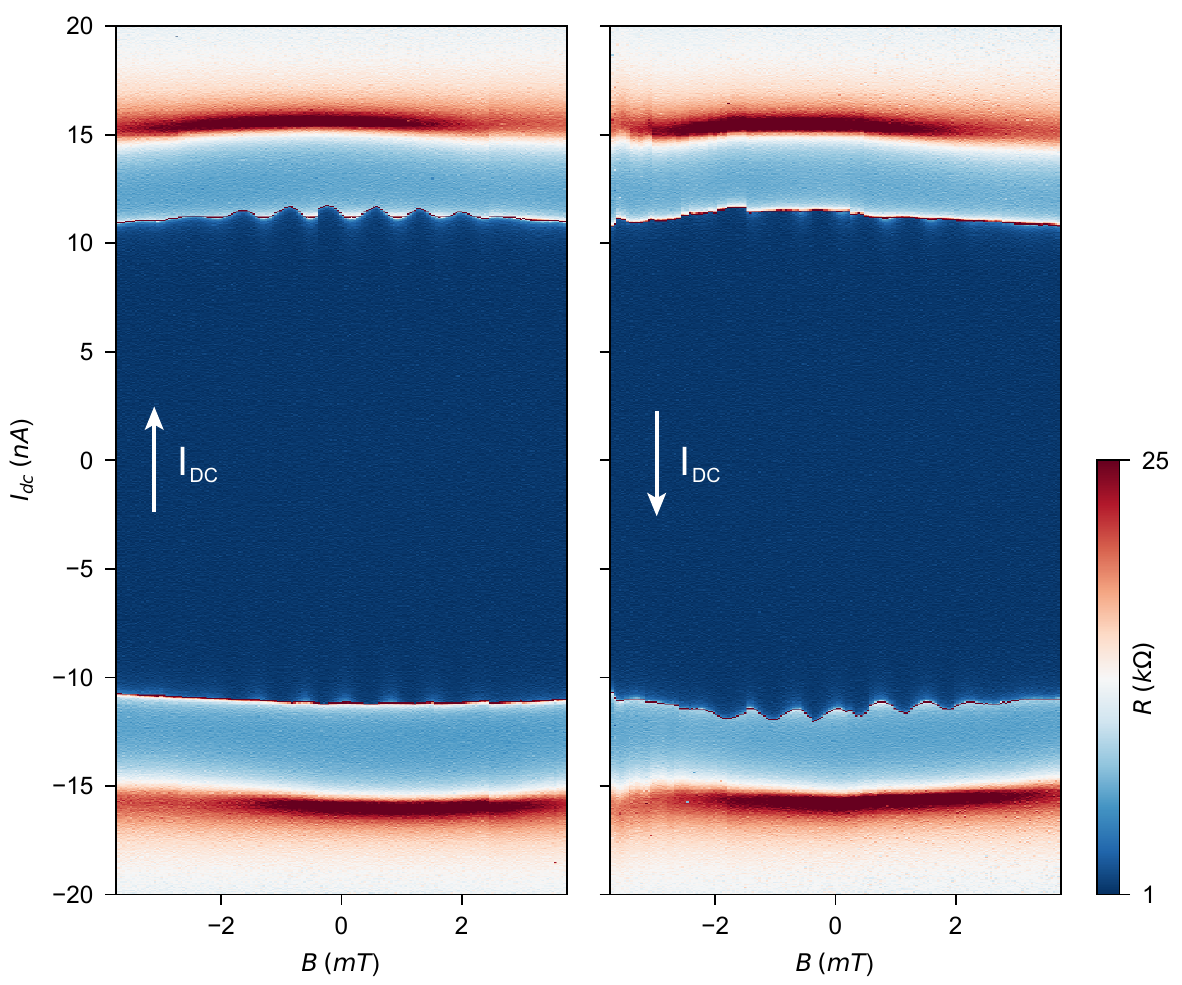}
\caption{\textbf{Presence or absence of oscillations depending on the current sweep direction.} The device is in the most asymmetric configuration. The left panel of the figure is a zoom in of Fig.~\ref{fig:1}. We measure the voltage drop across the device as a function of magnetic field and current. We observe oscillations due to superconducting interference in the superconducting lobe or the switching current but no oscillations for the retrapping current. The current bias is swept from negative to positive values in the left panel and from positive to negative in the right panel.}
\label{fig:S_A}
\end{figure*}

\clearpage
\newpage

\bibliographystyle{naturemag}
\bibliography{Bibliography}

\end{document}